\documentclass[pdftex,twocolumn,epjc3]{svjour3}          

\RequirePackage[T1]{fontenc}

\smartqed  

\RequirePackage{graphicx}
\RequirePackage{mathptmx}      
\usepackage{amsmath}
\RequirePackage{flushend}
\usepackage{booktabs}
\RequirePackage[colorlinks,citecolor=blue,urlcolor=blue,linkcolor=blue]{hyperref}

\newcommand{\nuc}[2]{$^{#1}$#2}
\newcommand{\expect}[1]{\left<#1\right>}
\renewcommand{\d}{\text{d}}
\newcommand{\asymmerror}[3]{$#1^{+#2}_{-#3}$}

\journalname{Eur. Phys. J. A}

\begin{document}

\title{Where we stand on structure dependence of ISGMR in the Zr-Mo region: Implications on $K_\infty$ }

\author{K. B. Howard \thanksref{addr1,e1}
        \and
        U. Garg \thanksref{addr1,e2}
        \and
        Y. K. Gupta \thanksref{addr2,e3}
        \and
        M. N. Harakeh \thanksref{addr3,e4}
}

\thankstext{e1}{e-mail: khoward5@nd.edu}
\thankstext{e2}{e-mail: garg@nd.edu}
\thankstext{e3}{e-mail: ykg.barc@gmail.com}
\thankstext{e4}{e-mail: m.n.harakeh@kvi.nl}

\institute{Department of Physics, University of Notre Dame, Notre Dame, Indiana 46556, USA \label{addr1}
          \and
          Nuclear Physics Division, Bhabha Atomic Research Centre, Mumbai 400085, India \label{addr2}
          \and
          KVI-CART, University of Groningen, 9747 AA Groningen, The Netherlands \label{addr3}
}

\date{Received: date / Accepted: date}

\maketitle

\begin{abstract}

Isoscalar giant resonances, being the archetypal forms of collective nuclear behavior, have been studied extensively for decades with the goal of constraining bulk nuclear properties of the equation of state, as well as for modeling dynamical behaviors within stellar environments.
An important such mode is the isoscalar electric giant monopole resonance (ISGMR) that can be understood as a radially symmetric density vibration within the saturated nuclear volume. The field has a few key open questions, which have been proposed and remain unresolved. One of the more prov-ocative questions is the extra high-energy strength in the $A\approx 90$ region, which manifested in large percentages of the $E0$ sum rule in \nuc{92}{Zr} and \nuc{92}{Mo} above the main ISGMR peak. The purpose of this article is to introduce these questions within the context of experimental investigations into the phenomena in the zirconium and molybdenum isotopic chains, and to address, via a discussion of previously published and preliminary results, the implications of recent experimental efforts on extraction of the nuclear incompressibility from this data.

\end{abstract}

\section{Background} \label{intro_general} Within the context of the scaling model as described in Ref. \cite{stringari_sum_rules}, one can calculate the nuclear incompressibility of a finite nucleus, $K_A$, from the energy of the compressional-mode electric isoscalar giant monopole resonance,
\begin{align}
  E_\text{ISGMR} &= \hbar \sqrt{\frac{K_A}{m\expect{r_0^2}} } \label{energy_KA},
\end{align}

\noindent where $m$ is the free-nucleon mass, and $\expect{r_0^2}$ is the ground-state mean-square nuclear mass radius. Generally, the ISGMR energies would be associated with one of the moment ratios $\sqrt{m_3/m_1}$, $m_1/m_0$, or $\sqrt{m_1/m_{-1}}$, where the moments $m_k$ of the strength function are defined generally as
\begin{align}
  m_k = \int S_\lambda(E_{x}) E_{x}^k \,  \d E_{x}, \label{moments_defined}
\end{align}
with $\lambda$ being the multipolarity of the resonance in question and $S_\lambda(E_x)$ being its associated strength distribution \cite{stringari_sum_rules}. Further, one should recall that Eq. \eqref{energy_KA}, relating $K_A$ with the resonance energies extracted in this manner, is predicated on the assumption that the strength distribution of the resonance is contained within a single collective peak \cite{harakeh_book}. Section \ref{results} contains a more complete description of these quantities.

It is well-established that measurements of $K_A$ in finite nuclei are the most direct means by which one can constrain the incompressibility of nuclear matter, $K_\infty$, defined as:
\begin{align}
K_\infty & = 9\rho_0^2 \frac{\d^2 \epsilon}{\d^2 \rho} \bigg|_{\rho=\rho_0}.
\end{align}

The nuclear incompressibility is thus a measure of the curvature of the nuclear equation of state, $\epsilon(\rho)$ at the saturation density of nuclear matter, $\rho_0$. $K_\infty$ is a bulk property of the nuclear force and thus should be invariant to the choice of the finite nucleus one uses to constrain its value. Indeed, this is the case, provided that approximately 100\% of the energy-weighted sum rule (EWSR) is exhausted within the peak of the ISGMR response \cite{harakeh_book}.

For details about how one obtains values of $K_\infty$ from finite nuclei, we refer the reader to Refs. \cite{blaizot,colo_2004a}; for further exposition on the ISGMR and for the models for extracting $K_A$ from experimental ISGMR strength distributions, Refs. \cite{stringari_sum_rules,stringari_lipparini_sum_rules,harakeh_book,garg_colo_review} are most comprehensive. It has been shown that the microscopic calculations of $K_\infty$ are strongly correlated with the ISGMR response of finite nuclei.
Thus, in a general sense, \emph{any} structure effects which are shown to influence the distribution of ISGMR strength would have substantial influence upon the calculated bulk properties of nuclear matter.

\begin{figure}[h]
\includegraphics[width=1\linewidth]{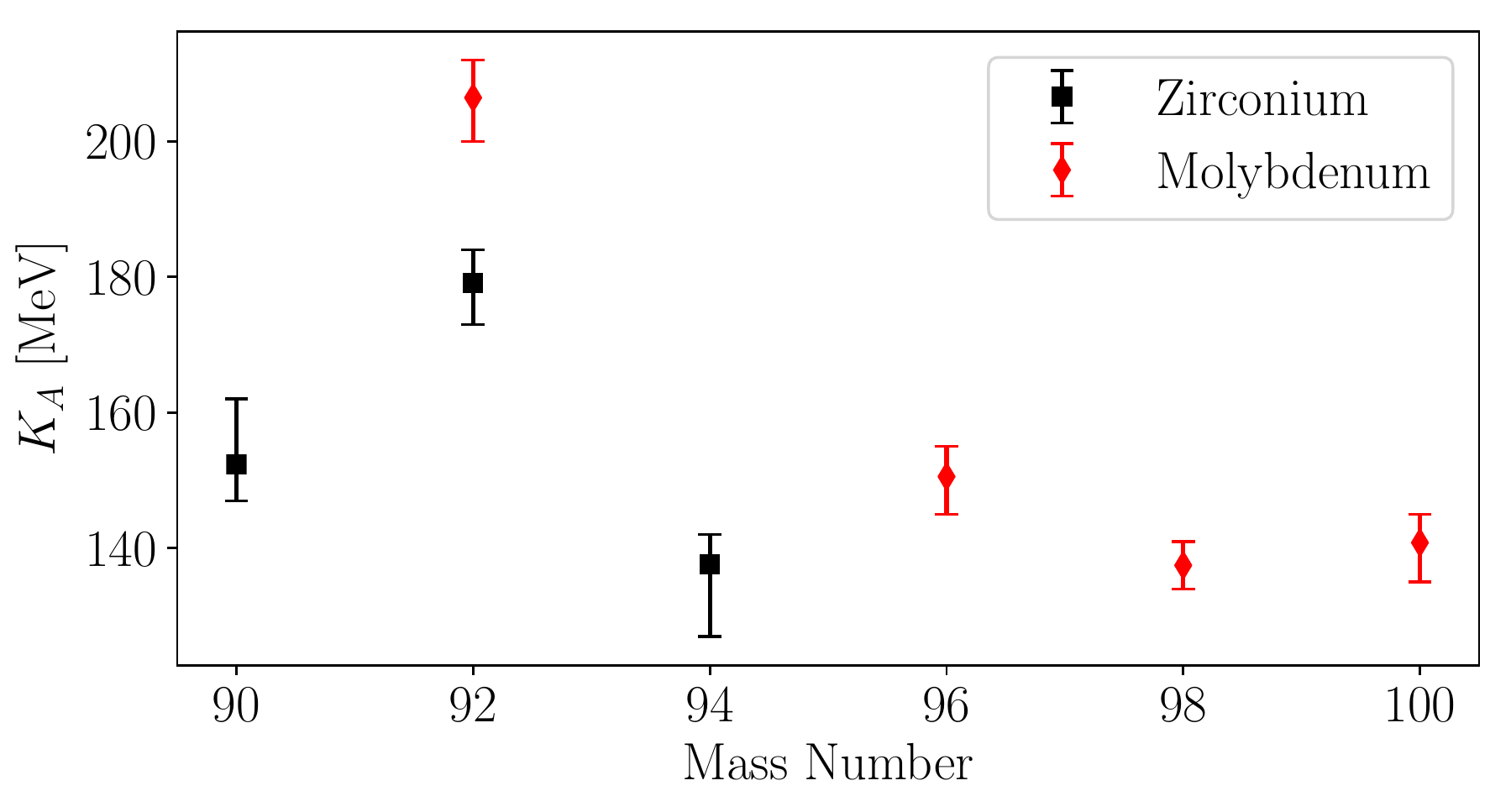}
\caption{(Color online) Experimental $K_A$ values extracted within the scaling model using the methodology of Refs.  \cite{youngblood_A90_unexpected,krishichayan_Zr,youngblood_92_100Mo,button_94Mo} for \nuc{90,92,94}{Zr} and \nuc{92,96,98,100}{Mo}. Shown is the reportedly stark disparity between extracted values of $K_A$ for the $A=92$ isobars relative to the other nuclei in this mass region. Data adapted from Ref. \cite{youngblood_A90_unexpected}.}
\label{AM_KA}
\end{figure}

As argued by the Texas A\&M (TAMU) group in Refs. \cite{youngblood_A90_unexpected,krishichayan_Zr,youngblood_92_100Mo,button_94Mo}, this conclusion has been challenged on the basis of experimental observations of the ISGMR strength in even-even isotopes of zirconium and molybdenum, namely, \nuc{90-94}{Zr} and \nuc{92-100}{Mo}. Figure \ref{AM_KA} illustrates these results. In particular, the results indicated that for \nuc{92}{Zr} and \nuc{92}{Mo}, a large portion of the $E0$ strength lies above the main ISGMR peak, resulting in $K_A$ values which are commensurately large. While the structure of the ISGMR in these nuclei is indeed important to the understanding of collective excitations, it should be kept in mind that, as previously stated, the association of $K_A$ with the GMR energies demands care, and can become untenable within the framework of Eq. \eqref{energy_KA} for multiply-peaked distributions of ISGMR strength.

A major primary motivation for studying the ISGMR is to probe bulk nuclear properties of the nuclear equation of state. As such, it is highly unexpected that effects arising from microscopic shell structure would appreciably influence the collective behavior of the nucleus undergoing these excitations.
As we shall discuss in the subsequent sections, the reported structure effects in the TAMU results themselves are in dispute, as results from our own independent experimental campaign into determining the nature of ISGMR strength for nuclei within this mass region seem to disagree with TAMU group's conclusions.

\section{The experiments} \label{experimental}

A pair of experiments were carried out at the Research Center for Nuclear Physics (RCNP), at Osaka University. Each used identical methodologies, one of the goals being to constrain the behavior of the ISGMR response in \nuc{90,92}{Zr} and \nuc{92}{Mo} within the context of the questions posed in Refs. \cite{youngblood_A90_unexpected,krishichayan_Zr,youngblood_92_100Mo,button_94Mo}. The present discussion will be restricted to the ISGMR data of \nuc{90,92}{Zr} and \nuc{92,94,96}{Mo}; a full description of the giant resonance strengths of \nuc{94-100}{Mo} will be presented in a forthcoming publication \cite{kevin-all}.

\begin{figure*}[t]
  \centering
\includegraphics[width=0.65\linewidth]{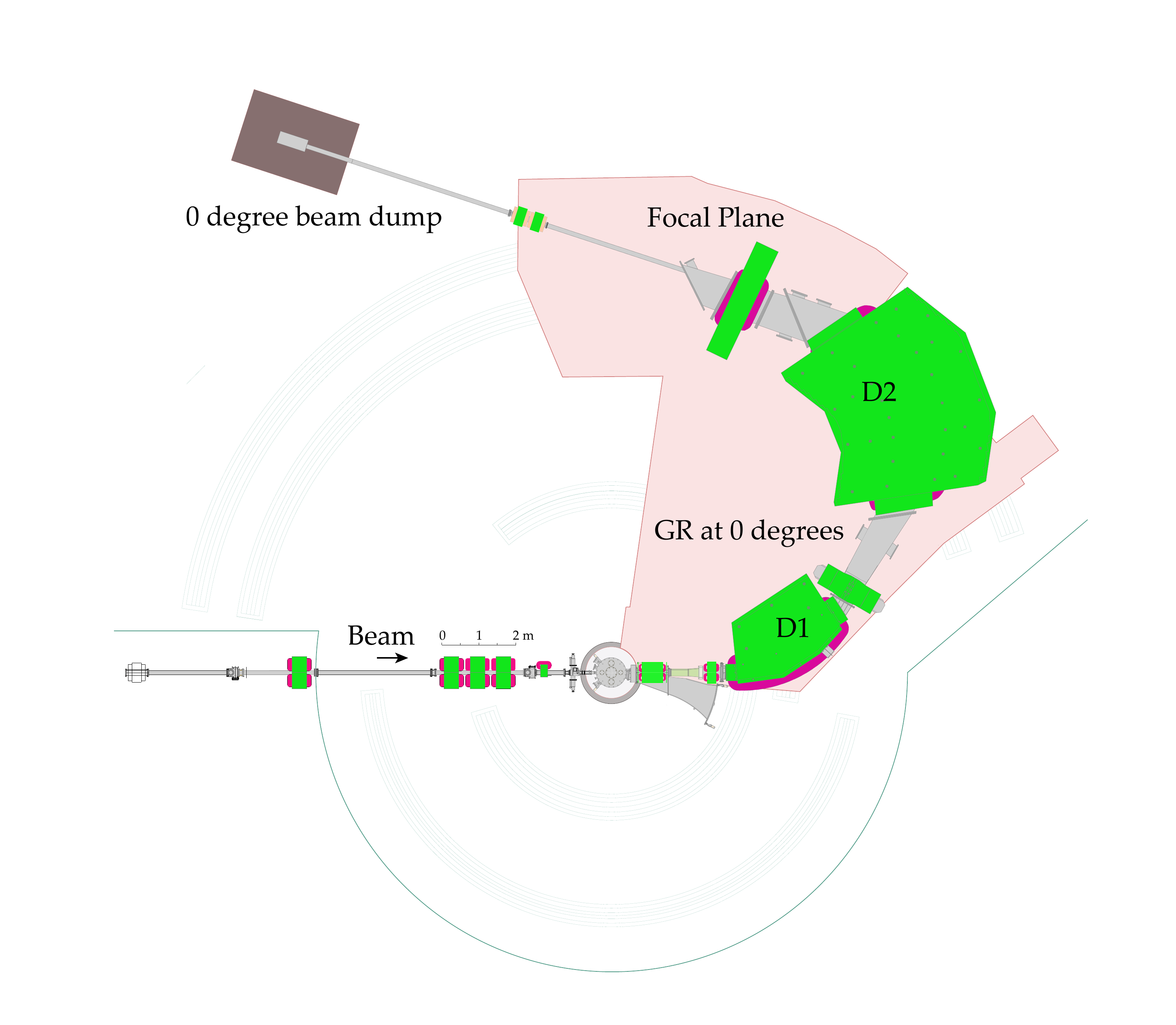}
\caption{(Color online) Schematic drawing of the Grand Raiden spectrometer in the zero-degree arrangement. Shown in green are the magnetic quadrupoles and dipoles; we have labeled the momentum-analyzing magnets D1 and D2. Figure courtesy of Prof. A. Tamii. Further details on the applicability of Grand Raiden to giant resonance studies can be found in Ref. \cite{itoh_sm_PRC}}
\label {GR}
\end{figure*}

In the experiments, $\alpha$-particles were accelerated by the coupled azimuthally-varying field and ring cyclotrons to a beam energy of $E_\alpha=386$ MeV. Zirconium and molybdenum targets with isotopic purity of approximately 95\% and areal densities of $\sim$5 mg/cm$^2$ were bombarded and the scattered $\alpha$-particles were then momentum analyzed by the high-precision mass spectrometer, Grand Raiden, a schematic drawing of which is presented in Fig. \ref{GR}. The focal-plane detector system was comprised of a pair of vertical and horizontal position-sensitive multiwire drift chambers in addition to plastic scintillators for particle identification. The vertical and horizontal positions at the focal plane allowed for a precise reconstruction of the scattering angles. The unreacted $\alpha$ beam passed unhindered at the high-energy side of the focal plane and was dumped in a well-shielded Faraday cup; see Fig. \ref{GR}.

\begin{figure*}[t]
\includegraphics[width=\linewidth]{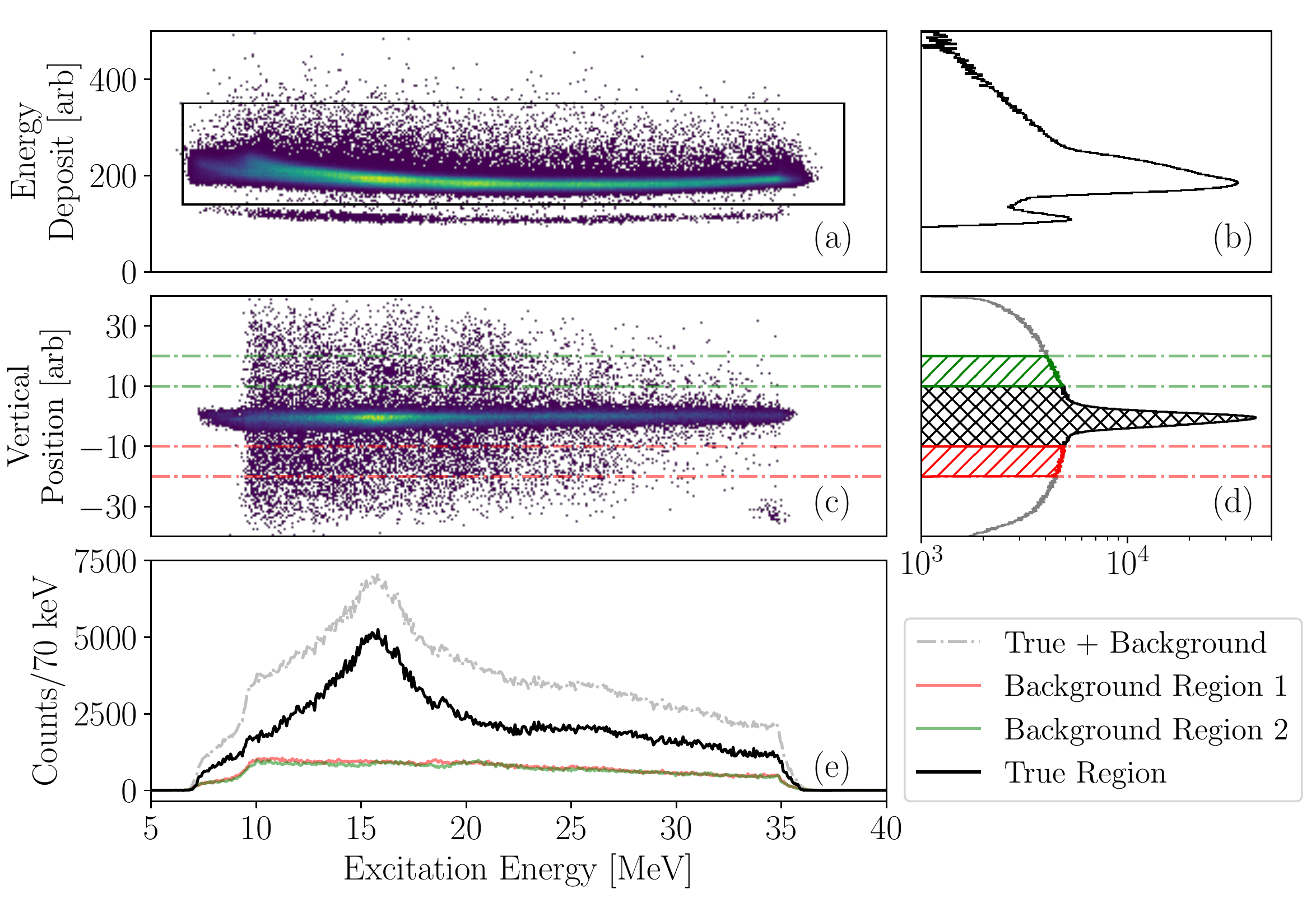}
\caption{(Color online) Depictions of the gates applied within the offline data reduction process in this work. \textbf{(a)} Particle identification spectrum, showing the energy deposited into a plastic scintillator against excitation energy. The enclosed, strong line shown corresponds to $\alpha$-events, which were gated upon in the offline analysis, while the excluded weaker line is comprised of events from other species which were rejected. \textbf{(b)} Projection of the scintillator energy deposition histogram onto the vertical axis. Visible is the strong $\alpha$-peak as well as the comparatively small rejected peak. \textbf{(c)} Two-dimensional histogram displaying the correlation between the energy-calibrated horizontal focal-plane position versus the vertical focal-plane position after application of the particle identification gate of (a). \textbf{(d)} Vertical focal-plane position of (c) projected onto the vertical axis. \textbf{(e)} Excitation-energy spectra for each of the hatched regions in (d), as well as the subtracted spectrum which is comprised essentially of instrumental-background-free $\alpha$-events.}
\label{background_excitation_spec}
\end{figure*}

Fig. \ref{background_excitation_spec} shows a series of plots which delineate the steps taken in the data reduction for these nuclei. The particle identification was completed via examination of the energy deposited into the scintillators located at focal plane. Figures \ref{background_excitation_spec}(a) and \ref{background_excitation_spec}(b) show, respectively, the correlation between energy-deposition and excitation energy as well as the one-dimensional energy-loss histogram. The enclosed region in (a) corresponds to $\alpha$-events which were gated upon in the offline analysis discussed hereafter, while the other events were rejected and correspond to other atomic species.

Figure \ref{background_excitation_spec}(c) and \ref{background_excitation_spec}(d) show typical vertical focal-plane position spectra. Operation of Grand Raiden in vertical focusing mode allows for true events which originate from scattering off the target to be coherently focused along the vertical plane, whereas events originating up- or down-stream from the scattering chamber (for example, from scattering off the beamline or collimator) are over- or under-focused. In Fig. \ref{background_excitation_spec}(d), the black doubly-hatched region corresponds to events which are focused to the median of the vertical focal-plane position and thus correspond to a combination of ``true'' events and those arising from instrumental background effects. The red and green singly-hatched regions correspond to gates on the off-median focal-plane positions in the spectra, which arise purely from instrumental background. This property of the measurement allows for a nearly complete and unambiguous subtraction of instrumental background.

The background contribution to the spectra is largest near forward angles, as the elastic cross sections are high and thus, elastically scattered particles which subsequently scatter off elements in the beamline can contribute to the background at this spectrometer setting. Further, we make the point that the various background gates shown in Fig. \ref{background_excitation_spec}(d) result in nearly identical background contributions to the excitation-energy spectra, as evidenced in Fig. \ref{background_excitation_spec}(e).

A precise energy calibration was obtained via examination of energy spectra from the \nuc{12}{C}$(\alpha,\alpha^\prime)$ and \nuc{24}{Mg}$(\alpha,\alpha^\prime)$ reactions, taken at each angular setting of Grand Raiden and at each magnetic field setting. The energy losses of the scattered $\alpha$-particles through the target foils are small, but were accounted for using SRIM calculations \cite{SRIM} under the assumption that the scattering event occurred at the midpoint of the foil. The acceptance of the spectrometer along the lateral dispersive plane ranged from excitation energies corresponding to approximately $10 \leq E_{x} \leq 32$ MeV. Angular distributions were extracted over a laboratory-frame angular range of $0^\circ - 10^\circ$ in each experiment. These central angles are then averaged over the acceptance of their solid angles and finally converted to the center-of-mass frame using the appropriate relativistic kinematics.

To constrain the optical model parameters (OMP) required for the analysis (Section \ref{data_analysis}), elastic scattering angular distributions as well as the cross sections of inelastic scattering to low-lying discrete states ($2_{1}^+$, $3_{1}^-$) were measured on \nuc{90,92}{Zr}, \nuc{92,98}{Mo}. These data were taken over an angular range of approximately $5^\circ-30^\circ$ in the laboratory frame.

\section{Data analysis}\label{data_analysis}

In order to reliably extract multipole strength distributions using the methodology presented here, it is necessary to have a reliable optical model with which one can perform calculations within the Distorted-Wave Born Approximation (DWBA) framework.

The computer code \texttt{PTOLEMY} was used for the DWBA calculations, using an optical model of the form:
\begin{align}
  U(r) &= \mathcal{V}_\text{Coul}(r) - \mathcal{V}_\text{vol} (r) - i \mathcal{W}_\text{vol}(r),
\end{align}
within which $\mathcal{V}_\text{Coul}$ is a point-sphere Coulomb potential, and $\mathcal{V}_\text{vol}$ and $\mathcal{W}_\text{vol}$ are chosen as the hybrid single-folding optical model prescribed by Satchler and Khoa \cite{khoa_satchler_single_folding}. In this model, the imaginary volume potential takes the shape of a Woods-Saxon function:
\begin{align}
  \mathcal{W}_\text{vol} (r) = \frac{W_\text{vol}}{1+\exp \left( \frac{r-R_I}{a_I}\right)},
\end{align}
while the real volume potential adopts the form of a point-nucleon Gaussian interaction which is folded with the target nuclear density and a modified density dependence:
\begin{align}
  \mathcal{V}_\text{vol} (r) = V_\text{vol} \int \d^3 \vec{r^\prime} \rho(r^\prime) f(\rho^\prime) \bar{v}_G(s).
\end{align}
Here, $V_\text{vol}$, $W_\text{vol}$, $R_I,$ and $a_I$ are free parameters in the optical model parameter (OMP) set found in the fitting procedure, while $s=|\vec{r}-\vec{r^\prime}|$ is the inter-particle distance, and
\begin{align}
  f(\rho^\prime) = 1-\zeta \rho^\beta (r^\prime) \notag \\
  \bar{v}_G (s) = \exp \left(s^2/t^2\right),
\end{align}
are the modified density dependence and Gaussian interaction. The parameters $\zeta = 1.9 \text{ fm}^2$, $ \beta = 2/3$, and $t = 1.88 \text{ fm}$ were adopted from Ref. \cite{khoa_satchler_single_folding}, along with the extension to the calculation of transition form-factors within this framework. The target nuclear densities, $\rho(r^\prime)$, are taken to be two-parameter Fermi distributions and are available from Ref. \cite{fricke_ADNDT}.

\begin{table*}[]
\centering
  \begin{tabular}{@{}ccccccccccccc@{}}
  \toprule
               & \multicolumn{4}{c}{Optical Model Parameters}                          &  & \multicolumn{2}{c}{Density Parameters} &  & \multicolumn{2}{c}{$2_{1}^+$}                         & \multicolumn{2}{c}{$3_{1}^-$}                         \\ \cmidrule(lr){2-5} \cmidrule(lr){7-8} \cmidrule(l){10-13}
  Nucleus      & $V_\text{vol}$  & $W_\text{vol}$ & $R_I$  & $a_I$  &  & $c$            & $a$           &  & $E_{x}$  & $B(E2)$  & $E_{x}$  & $B(E3)$  \\
        & [MeV]  &  [MeV] &  [fm] & [fm] &  & [fm]           &  [fm]          &  &[MeV] &  [$\text{e}^{2} \text{b}^{2}$] &  [MeV] &  [$\text{e}^{2} \text{b}^{3}$] \\ \cmidrule(r){1-5} \cmidrule(lr){7-8} \cmidrule(l){10-13}
  \nuc{90}{Zr} & 37.6                 & 35.5                 & 6.13       & 0.623      &  & 4.908              & 0.523             &  & 2.186         & 0.061                                 & 2.740         & 0.056                                 \\ \cmidrule(r){1-5} \cmidrule(lr){7-8} \cmidrule(l){10-13}
  \nuc{92}{Zr} & 35.4                 & 38.8                 & 6.02       & 0.687      &  & 4.958              & 0.523             &  & 0.934         & 0.083                                 & 2.339         & 0.075                                 \\ \cmidrule(r){1-5} \cmidrule(lr){7-8} \cmidrule(l){10-13}
  \nuc{92}{Mo} & 32.4                 & 40.4                 & 6.04       & 0.610      &  & 4.975              & 0.523             &  & 1.509         & 0.097                                 & 2.849         & 0.077                                 \\ \cmidrule(r){1-5} \cmidrule(lr){7-8} \cmidrule(l){10-13}
  \nuc{98}{Mo} & 30.5                 & 47.2                 & 5.19       & 1.090      &  & 5.105              & 0.523             &  & 0.787         & 0.267                                 & 2.017         & 0.133                                 \\ \bottomrule
  \end{tabular}
\caption{Table listing the optical-model parameters extracted from fits to elastic scattering angular distributions and used for the DWBA input to the multipole-decomposition analyses. The efficacy of these OMPs are shown in Fig. \ref{elastics}. Also shown are the low-lying $2_{1}^+$ and $3_1^-$ excitations in addition to the reduced transition probabilities from Refs. \cite{raman_ADNDT,kibedi_ADNDT}.}
\label{OMP_table}
\end{table*}

\begin{figure*}[]
  \includegraphics[width=1.0\linewidth]{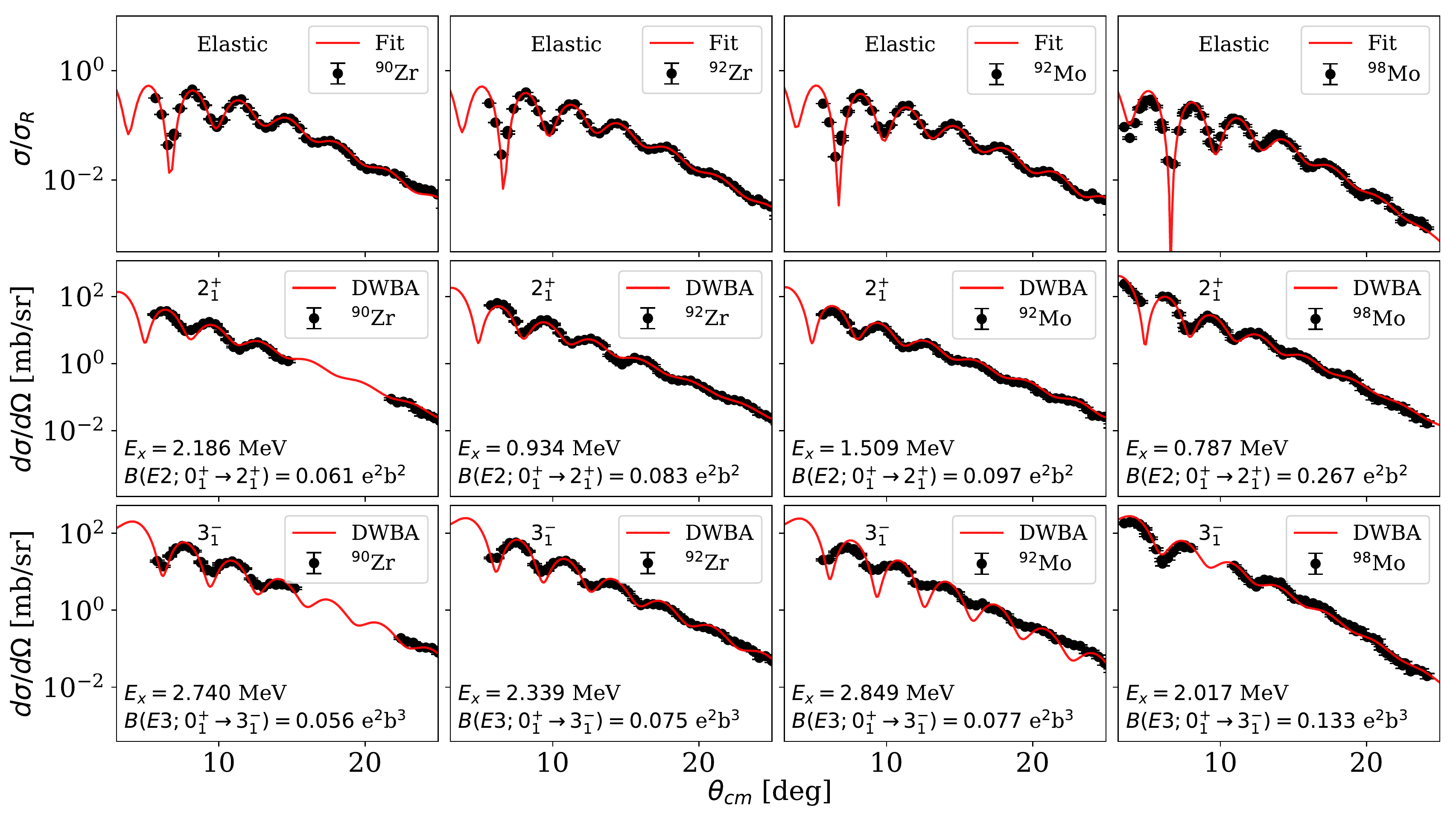}
  \caption{(Color online) Top: Elastic scattering angular distributions for each nucleus, normalized to the Rutherford cross section, shown with results of the optical-model fits obtained with parameters listed in Table \ref{OMP_table}. Middle: Angular distribution of differential cross section for excitation of the $J^\pi = 2_1^+$ state, and results of DWBA calculations using the optical-model parameters obtained from fitting the elastic scattering data, with adopted $B(E2)$ values from Ref. \cite{raman_ADNDT}. Bottom: Same as above, but for the $J^\pi = 3^-_1$ state, with $B(E3)$ values from Ref. \cite{kibedi_ADNDT}. }
  \label{elastics}
\end{figure*}

Results of the least-$\chi^2$ analysis for the elastic scattering angular distributions within this model framework are shown in Fig. \ref{elastics}, with the resulting OMPs listed in Table \ref{OMP_table}. Validation of the predictive power of an OMP-set, and its ability to reproduce inelastic scattering angular distributions was tested on the experimentally available low-lying discrete states. The angular distributions of inelastic scattering to the $2_1^+$ and $3_1^-$ states were calculated in the DWBA framework using the previously-known $B(E\lambda)$s.
These are compared in Fig. \ref{elastics} with the experimental angular distributions and show excellent agreement.

The inelastic scattering spectra were sorted into 1 MeV-wide bins for \nuc{90,92}{Zr} and \nuc{92}{Mo} (500 keV for \nuc{94,96}{Mo}) at each angle and angular distributions were extracted for each excitation energy. Using the OMPs from elastic scattering data (see Table \ref{OMP_table}), a multipole-decomposition analysis (MDA) was carried out whereby the experimental angular distributions are decomposed into a superposition of angular distributions corresponding to pure angular momentum transfers of $\lambda = 0$ to $\lambda = 10$. For \nuc{94,96}{Mo}, elastic scattering data were not measured, and the optical model parameters obtained from \nuc{98}{Mo} within the same experiment were instead utilized for the subsequent multipole decomposition. The MDA is defined as follows:

\begin{align}
  \frac{\d^2\sigma^\text{exp}(\theta_\text{c.m.},E_x)}{\d \Omega\, \d E} & = \sum_\lambda A_\lambda (E_x) \frac{\d^2\sigma^\text{DWBA}_\lambda(\theta_\text{c.m.},E_x)}{\d \Omega \, \d E}.
\end{align}

If the DWBA calculations are completed using coupling parameters which correspond to $100\%$ of the EWSR, then $A_\lambda$ corresponds to the fraction of the corresponding EWSR exhausted within that particular energy bin \cite{harakeh_book,satchler_isospin,Li_PRL,Li_PRC}. The distributions of isovector giant dipole resonance (IVGDR) strength for these nuclei are known from Refs. \cite{berman_GDR,plujko_GDR}, and those, in combination with DWBA calculations incorporating the Goldhaber-Teller model \cite{satchler_isospin}, allow for the IVGDR strengths to be explicitly accounted for in the MDA procedure. Although multipolarities were included up to $\lambda_\text{max} = 10$, our angular range is sufficient to reliably extract strengths only for $\lambda \leq 2$; the extracted monopole strengths are insensitive to increasing values of $\lambda_\text{max}$, however.

\begin{figure*}[h!]
  \includegraphics[width=\linewidth]{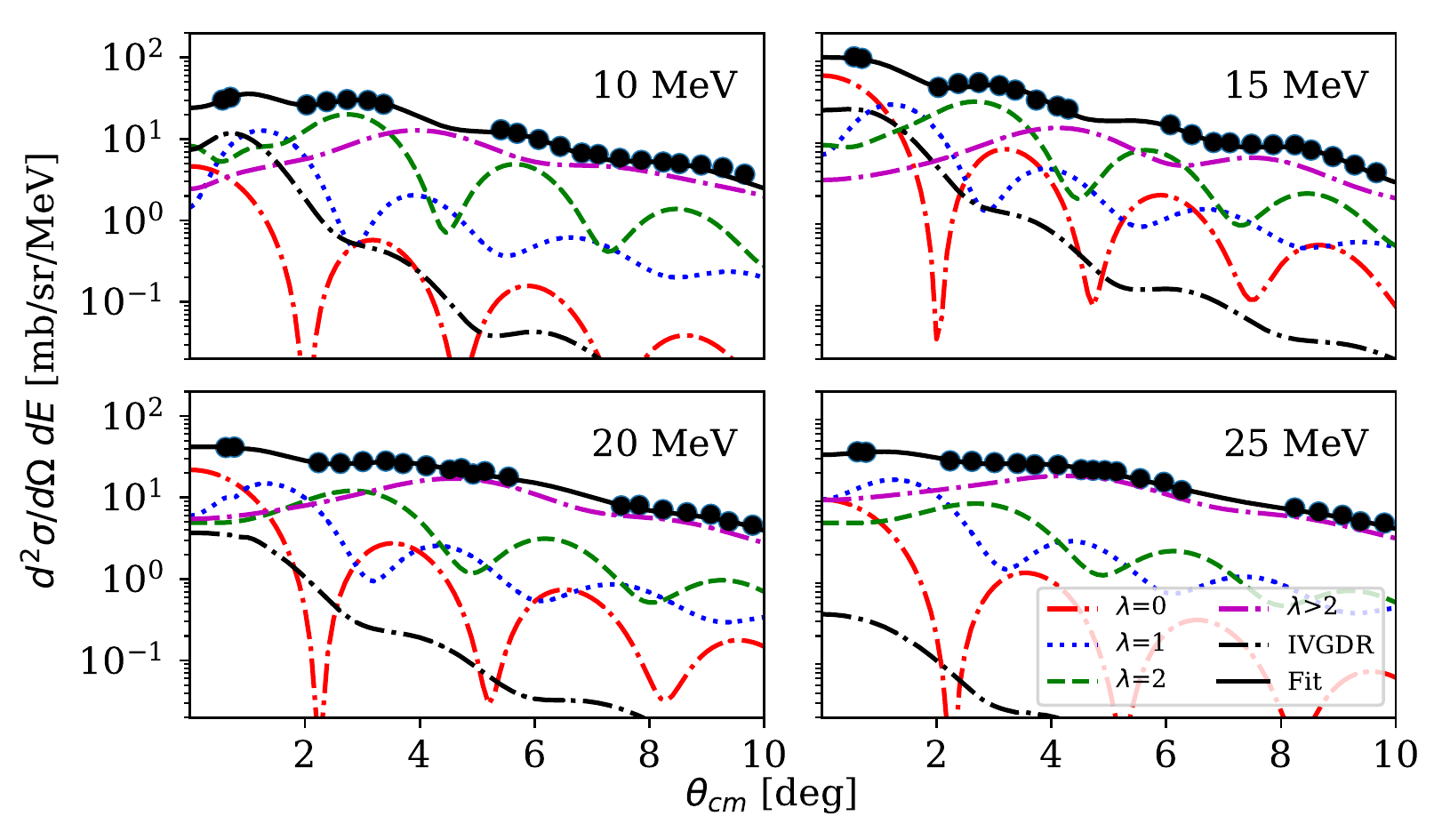}
  \caption{(Color online) Results of the MDA for \nuc{94}{Mo} for excitation-energy bins centered at $10$, $15$, $20$ and $25$ MeV. Shown are $\lambda = 0,1,2$, as well as the contribution from the IVGDR, other multipoles $\lambda>2$, and the total fit distribution to experimental data. Gaps in the data correspond to angular regions where the contribution by the elastic scattering channel from the hydrogen contamination present in the targets is dominant. These results are typical for various energies and for all nuclei present in the study.  }
  \label{MDA_94Mo}
\end{figure*}

The details of the MDA procedure employed here, as well as its suitability with regards to estimations of parameter uncertainties, are discussed in Ref. \cite{gupta_A90_PRC}. The Python implementation, \texttt{emcee}, for the Markov-Chain Monte-Carlo algorithm of Goodman and Weare was employed \cite{foreman_mackey,goodman_weare}. The algorithm allows for the generation of multidimensional probability distributions for $A_\lambda$ coefficients. The $68\%$ confidence interval, centered at the distribution mean, was taken as the uncertainty in each parameter.

Shown in Fig. \ref{MDA_94Mo} is a subset of the multipole decompositions obtained in the analysis of \nuc{94}{Mo}. The figures are largely representative of the results of the MDA for all nuclei.

\section{Results}\label{results}
\begin{figure}[h!]
  \includegraphics[width=\linewidth]{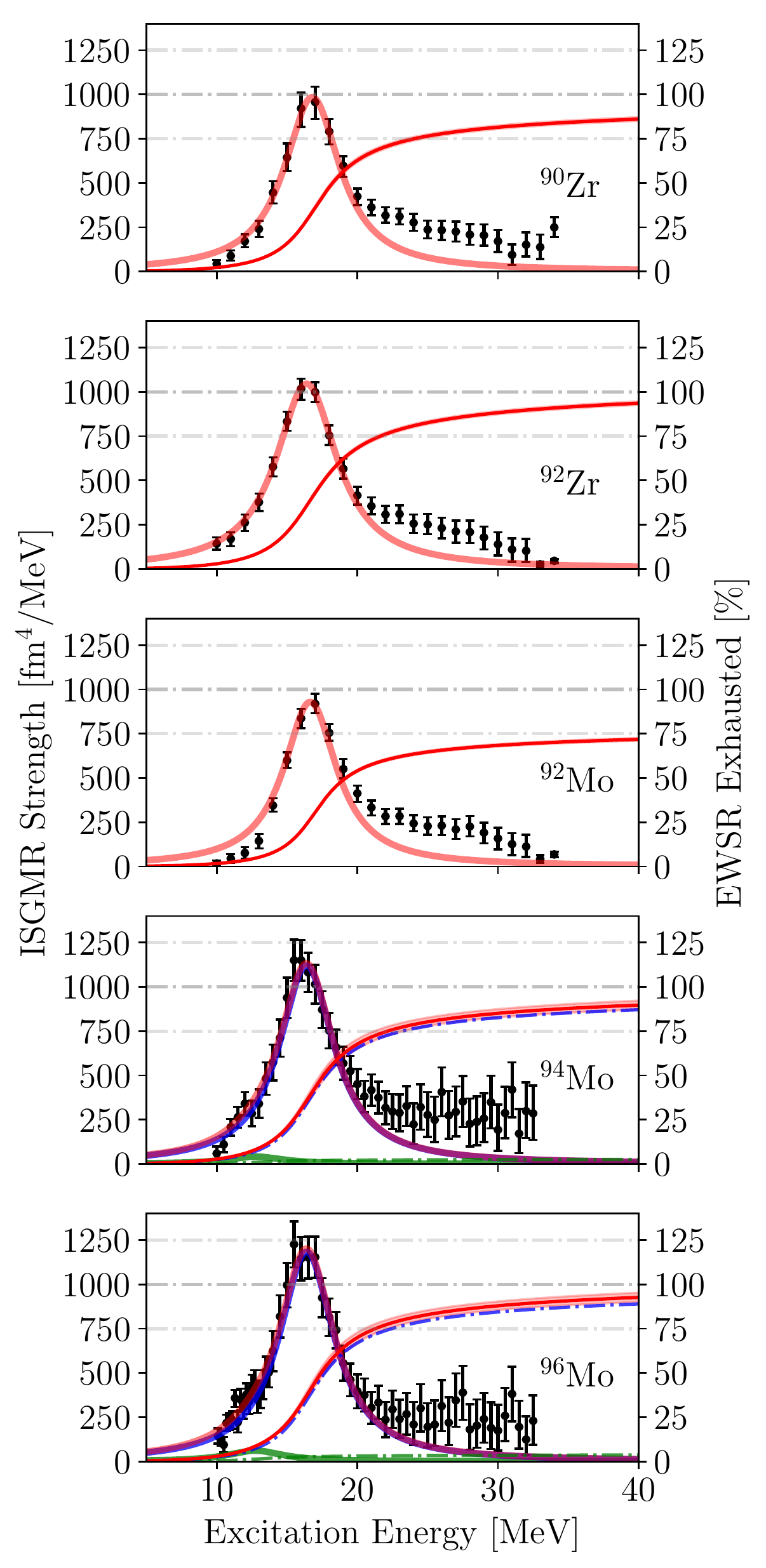}
  \caption{(Color online) Extracted ISGMR strengths for \nuc{90,92}{Zr} and \nuc{92,94,96}{Mo}, along with the fitted Lorentzian distributions from Table \ref{fit_parameters}. Shown on the right axes is the cumulative distribution function, or integrated EWSR, which has been identified in the fitted peaks, with the shaded region indicating the propagated uncertainties to the cumulative EWSR. Uncertainties arising due to the model-dependence of the optical model ($\sim 20\%$ of the EWSR magnitude) are not shown. In \nuc{92}{Mo}, a different choice of optical model could increase the EWSR exhausted. Published data from Ref. \cite{gupta_A90_PRC}.}
  \label{monopole_strengths_with_CDFs}
\end{figure}

The extracted ISGMR strengths are shown in Fig. \ref{monopole_strengths_with_CDFs}, in addition to the Lorentzian distributions, which were fitted to the data:

\begin{align}
    S(E_x,S_0,E_{0},\Gamma) & = \frac{S_0 \Gamma}{\left(E_x - E_0\right)^2 + \Gamma^2}.
    \label{lorentz}
\end{align}

Figure \ref{monopole_strengths_with_CDFs} also shows the running EWSR exhausted by the obtained Lorentzian distributions.  In further analyses of \nuc{94-100}{Mo}, it was found that deformation effects became manifest in the more neutron-rich nuclei. To account for this, the ISGMR strength distributions for those nuclei were fitted with a constrained combination of two peaks to account for potential coupling of the ISGMR strength with the $K=0$ component of the ISGQR \cite{garg_sm_PRL,kvi-def,itoh_sm_PRC,gupta_24Mg_plb,gupta_24Mg_prc,peach_28Si_prc}.

\begin{table*}[]
  \centering
\begin{tabular}{@{}ccccccccccc@{}}
\toprule
             &  & \multicolumn{3}{c}{Low Peak}                       &  & \multicolumn{3}{c}{High Peak} & & Total $E0$           \\ \cmidrule(lr){3-5} \cmidrule(l){7-9}
Nucleus      &  & $E_0$       & $\Gamma$   & EWSR             &  & $E_0$       & $\Gamma$ & EWSR   & & Assigned EWSR \\

      &  & [MeV]      &  [MeV]  &  [\%]            &  &  [MeV]      & [MeV]  & [\%]  & & [\%]\\
\cmidrule(r){1-5} \cmidrule(l){7-9} \cmidrule(l){11-11}
\nuc{90}{Zr} &  & -                & -               & -                    &  & $16.8 \pm 0.2$ & $2.4 \pm 0.4$ & $84 \pm 2$ & & $84\pm2$ \\ \cmidrule(r){1-5} \cmidrule(l){7-9} \cmidrule(l){11-11}
\nuc{92}{Zr} &  & -                & -               & -                    &  & $16.4 \pm 0.1$ & $2.2 \pm 0.3$ & $91 \pm 2$ & & $91 \pm 2$ \\ \cmidrule(r){1-5} \cmidrule(l){7-9} \cmidrule(l){11-11}
\nuc{92}{Mo} &  & -                & -               & -                    &  & $16.5 \pm 0.1$ & $2.3 \pm 0.1$ & $73 \pm 2$ & & $73 \pm 2$\\ \cmidrule(r){1-5} \cmidrule(l){7-9} \cmidrule(l){11-11}
\nuc{94}{Mo} &  & $12.7 \pm 0.5$ & $2.4 \pm 0.4$ & \asymmerror{2}{3}{2} &  & $16.4 \pm 0.21$ & $2.4 \pm 0.4$  & $86 \pm 3$ & & $88 \pm 4$\\ \cmidrule(r){1-5} \cmidrule(l){7-9} \cmidrule(l){11-11}
\nuc{96}{Mo} &  & $12.7 \pm 0.5$  & $2.3 \pm 0.3$ & \asymmerror{4}{3}{4} &  & $16.4 \pm 0.2$ & $2.4 \pm 0.3$ & $89 \pm 3$  & & $93 \pm 4$\\ \bottomrule
\end{tabular}
\caption{Fit parameters for each nucleus in the two experiments. Data are fit to one- or two-peak Lorentzian distributions (Eq. \eqref{lorentz}). Listed also is the integrated EWSR underneath the fitted peaks up to an excitation energy of $35$ MeV.}
\label{fit_parameters}
\end{table*}

In the analysis of \nuc{90,92}{Zr} and \nuc{92}{Mo} in the earlier measurements \cite{gupta_A90_PLB,gupta_A90_PRC}, only one peak was found sufficient for the description of the ISGMR response. The parameters for the Lorentzian-distribution fits to the experimental ISGMR strength distributions are presented in Table \ref{fit_parameters}.
In the cases of \nuc{94,96}{Mo}, although a second peak was included in the modeling of the data, the extracted EWSR for the low-energy peak is consistent with $0\%$. This would suggest that the deformation effects (and thus, any shifting of the ``main'' ISGMR peak) are negligible insofar as a comparison with the peak energies of \nuc{90,92}{Zr}, \nuc{92}{Mo} data is concerned. The uncertainties in the parameters shown in Table \ref{fit_parameters} are somewhat higher for \nuc{94,96}{Mo} due to the inclusion of a second, highly-correlated peak in the fitting procedure, but are still consistent with the results of \nuc{90,92}{Zr} and \nuc{92}{Mo}. Hereafter, we will refer only to the main ISGMR peak in the discussion.

We report that the peaks appear in the same location within the experimental-fit uncertainties. Even further, \nuc{92}{Zr} and \nuc{92}{Mo} are characterized by nearly identical locations of the ISGMR response, as determined from the fitting procedure, with a complete absence of any coherent peaks in the strength distribution above 20 MeV.

The distribution of strength extracted over the energy range $10 \leq E_{x} \leq 35$ MeV can be characterized by various moment ratios \cite{harakeh_book,stringari_sum_rules,jennings_jackson_constrained_scaling}:

\begin{align}
  E_\text{constrained} &= \sqrt{\frac{m_1}{m_{-1}}} \notag \\
  E_\text{centroid} & = \frac{m_1}{m_0} \notag \\
  E_\text{scaling} &= \sqrt{\frac{m_3}{m_1}}. \label{moment_ratios}
\end{align}

\begin{figure}[h!]
  \includegraphics[width=1.0\linewidth]{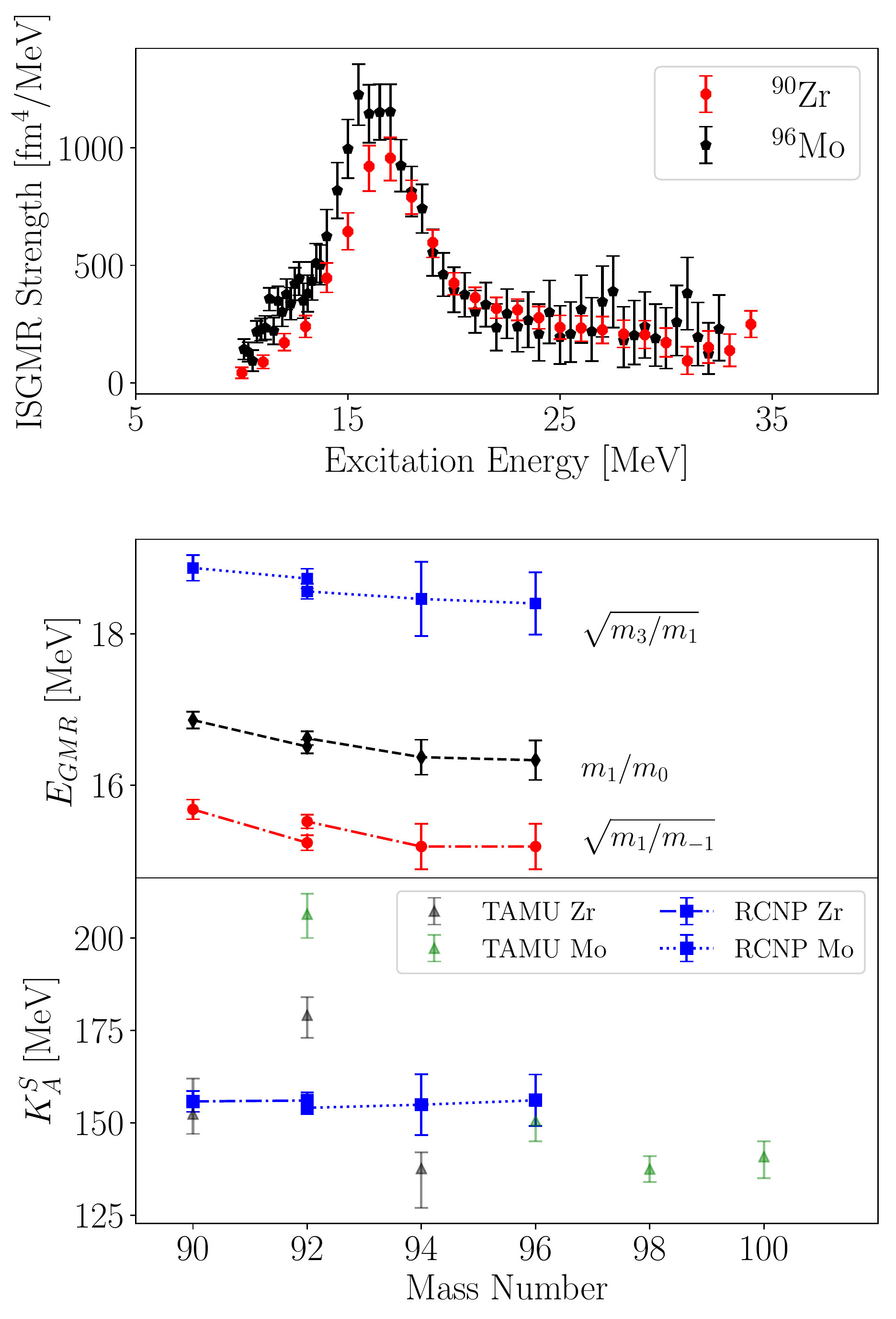}
  \caption{(Color online) Top: ISGMR strength distributions of the nuclei at the mass-extrema in this study: \nuc{90}{Zr} and \nuc{96}{Mo}. Evident is the structural and positional agreement of the distributions. Middle: Various moment ratios for the nuclei in this study. Lines connect \nuc{90,92}{Zr} and \nuc{92-96}{Mo}. Bottom: TAMU extractions of $K_A^S$ shown previously in Fig. \ref{AM_KA} from Refs. \cite{youngblood_A90_unexpected,krishichayan_Zr,youngblood_92_100Mo,button_94Mo} (black and green triangles), juxtaposed with the finite nuclear incompressibilities $K_A^S$ (blue squares) measured for \nuc{90,92}{Zr} and \nuc{92-96}{Mo} within these works. Shown clearly is a near-constant scaling-model nuclear incompressibility for the nuclei in this mass region.  In all cases, we have calculated $K_A$ from the resonance energies of Eq. \eqref{moment_ratios} over an energy range within which we have identified nearly $100\%$ of the EWSR, using the resonance energies listed in Table \ref{moment_ratios_total_EWSRS}. }
  \label{incompressibilities}
\end{figure}

These moment ratios were calculated from the extracted ISGMR peaks and are listed in Table \ref{moment_ratios_total_EWSRS}. Within the extracted
uncertainties, it is evident that the results for $^{90,92}$Zr and $^{92-96}$Mo for any given moment ratio are largely
in agreement with one another. This is shown graphically in Fig. \ref{incompressibilities}. Further, the resonance energies associated with the various moment ratios generally obey very well the general empirical trend of $E_\text{GMR} \sim A^{-1/3}$.



\begin{table}[]
\centering
\begin{tabular}{@{}cccc@{}}
\toprule
Nucleus      & $\sqrt{m_{1}/m_{-1}}$ & $m_1/m_0$   & $\sqrt{m_{3}/m_{1}}$   \\
     & [MeV] & [MeV]  &  [MeV]  \\ \midrule
\nuc{90}{Zr} & $15.7 \pm 0.1$            & $16.9 \pm 0.1$ & $18.9 \pm 0.2$          \\
\nuc{92}{Zr} & $15.2 \pm 0.1$            & $16.5 \pm 0.1$ & $18.7 \pm 0.1$            \\
\nuc{92}{Mo} & $15.5 \pm 0.1$            & $16.6 \pm 0.1$ & $18.6 \pm 0.1$           \\
\nuc{94}{Mo} & $15.2 \pm 0.3$            & $16.4 \pm 0.2$ & $18.5 \pm 0.5$           \\
\nuc{96}{Mo} & $15.2 \pm 0.3$            & $16.3 \pm 0.2$ & $18.4 \pm 0.4$            \\ \bottomrule
\end{tabular}
\caption{Moment ratios of Eq. \eqref{moment_ratios} calculated up to excitation energy $35$ MeV from the fit distributions of Table \ref{fit_parameters}.}
\label{moment_ratios_total_EWSRS}
\end{table}

$E_\text{constrained}$ and $E_\text{scaling}$ can be respectively associated with finite incompressibilities $K_A^C$ and $K_A^S$.
Within the constrained model, there is a radial dependence of the nuclear density, whereas in the scaling model the nuclear density changes uniformly \cite{jennings_jackson_constrained_scaling}. These finite nuclear incompressibilities can be calculated using the empirical density distributions parameterized in Table \ref{OMP_table}, Eq. \eqref{energy_KA}, and the appropriate moment ratio.


It is clear that the previously reported enhanced nuclear incompressibility of the $A=92$ isobars is not observed in the present work. The additional ISGMR peak reported in Refs. \cite{youngblood_A90_unexpected,krishichayan_Zr,youngblood_92_100Mo,button_94Mo} is not present in the results of our analysis, which provides the justification for modeling the peak that appears within the giant resonance region with Eq. \eqref{lorentz}. As this peak is shown generally to exhaust $\approx 100\%$ of the EWSR over the excitation-energy range of our experiments, our calculation of $K_A$ from the modeled line-shape may be deemed as valid. The obvious question remains as to what caused this difference in extracted strength above $20$ MeV which is reported by the TAMU group. It was argued in Refs. \cite{gupta_A90_PLB,gupta_A90_PRC} that the modeling of the instrumental background could introduce some effects of this nature. Some studies have been done to ascertain the sensitivities of the giant resonance strengths to the choice of continuum in this alternative method of analysis \cite{youngblood_24Mg_continuum_study}. In any event, as the background-subtraction described in Section \ref{experimental} and graphically depicted in Fig. \ref{background_excitation_spec} endeavors to measure the instrumental background itself and to subtract it from the data prior to analysis with no assumptions or arbitrariness, it may be concluded that our methodology for isolating the ISGMR response is more reliable.

Inspection of Fig. \ref{incompressibilities} indicates that the nuclear incompressibilities of \nuc{90,92}{Zr} and \nuc{92-96}{Mo} are consistent within the scaling model. In addition to answering the question of the enhanced nuclear incompressibility of \nuc{92}{Zr} and \nuc{92}{Mo}, one concludes immediately that \nuc{96}{Mo} is exactly as incompressible as \nuc{90}{Zr}, which is one of the ``standard'' nuclei to which many theoretical models are benchmarked. Further, the results for the ISGMR energy of \nuc{90}{Zr} are very well-consistent with the results of Ref. \cite{krishichayan_Zr}.

\section{Conclusions}
It was previously argued that the current understanding of the collective-model description of the giant resonance allows for the determination of the nuclear incompressibility only if the detailed effects of nuclear structure do not play a role in the positioning of the ISGMR energy \cite{gupta_A90_PLB}. The structure effects that reportedly manifested in \nuc{92}{Zr} and \nuc{92}{Mo} are disputed in the results of the present work reporting an independent measurement of the ISGMR strength distributions in this mass region.

It is clear from the moment ratios and extracted scaling-model incompressibilities of Fig. \ref{incompressibilities} that there does not seem to be any major differences manifesting along the zirconium or molybdenum isotopic chains. This is seen even more plainly from inspection of the strength distributions themselves, as also shown in Fig. \ref{incompressibilities}. The ISGMR strength of nuclei between \nuc{90}{Zr} and \nuc{96}{Mo} structurally look nearly identical; any differences could be easily explained within models depicting the resonance energy scaling inversely with the nuclear radius.

\begin{acknowledgements}
This work was supported in part by the National Science Foundation Grant No. PHY-1713857. KBH gratefully acknowledges support from the Arthur J. Schmitt Foundation, from the Liu Institute for Asia and Asian Studies, University of Notre Dame, and from the Japan Society for the Promotion of Science.
\end{acknowledgements}

\bibliographystyle{spphys}       
\bibliography{A90_epja}   

\end{document}